\documentclass[conference]{IEEEtran}

\usepackage{cite}
\usepackage{comment}
\usepackage[utf8]{inputenc}
\usepackage{dirtytalk}

\usepackage{graphicx}
\graphicspath{{Images/}}
\usepackage{multirow}
\usepackage{amsmath}
\usepackage{breqn}
\usepackage{float}

\ifCLASSINFOpdf
\else
\fi
\begin{document}
%
\title{Coordination of Transmission and Distribution Systems in Load Restoration}

\author{\IEEEauthorblockN{Santosh Sharma}
\IEEEauthorblockA{Department of Electrical \& Computer Engineering\\
University of Central Florida, Orlando, FL, USA, 32816}
}


%


\maketitle

\begin{abstract}
The power distribution system is evolving in the form of smart grid. The proliferation of distributed energy resources (DERs) is making the previously passive system active and more complicated. With the adoption of de-carbonization principles, large-scale coal and nuclear power plants are being gradually replaced by renewables and carbon-free DERs. With this rapid transformation, power system is operating with less inertia and minimal margins. In recent years, power system is facing apocalyptic weather events more frequently, and large-scale blackout have become regulars. After the complete or partial blackouts, power system goes through different stages before it reaches the normal operating condition. The load restoration is the stage where power system is fully established after the blackouts; however, due to the limiting ramping rates of centralized generation, the energization of large amounts of loads is delayed by some time. In order to mitigate the negative impact of ramping rates of centralized generation, DERs in distribution systems are proposed to serve the loads in both transmission and distribution systems in coordination with limited centralized generation in transmission system. The problem is formulated as centralized or integrated transmission and distribution (T$\&$D) coordination model. The modified IEEE 14 bus test case and IEEE 13 node test feeders are used to validate the proposed strategy; the results indicate the validity of the proposed model. 
\end{abstract}


%
\IEEEpeerreviewmaketitle

\section{Introduction}
In recent years, high-impact low frequency (HILF) events such as hurricanes, ice storms, earthquakes, cyber attacks, etc. are happening at relatively high frequency \cite{hines2008trends}. Impact of such HILF events is colossal, and it is reported that such events cost billions of dollars to United States every year \cite{noaa2018us}. Frequent happening of such disastrous events has highlighted the importance of resilience of critical infrastructure (CI) systems such as electricity, water delivery, communication systems, health, finance etc \cite{10415053,SHARMA2024122588}. Among all, power system is the most critical infrastructure system because it provides functional facilities to all other CI systems. In this context, power system resilience has been recognized as promising solution to enhancement of resilience of all CI systems. Power system resilience is defined as \say{the ability to prepare for and adapt to changing conditions and withstand and recover rapidly from extreme outages} \cite{che2018adaptive}. Therefore, to continue operating in case of disastrous events, power system resilience should be enhanced in several different ways at different stages of disastrous events. Recent research efforts related to power system recovery to enhance power system resilience will be discussed first in following few paragraphs.

Efficient pre-disaster planning and scheduling of emergency response resources play vital role in reducing impact of hazardous events and in effective service restoration after the disasters. Repair crews and emergency mobile power supplies are examples of emergency response resources. To reduce the impact of outages on customers, reference \cite{lei2016mobile} has investigated pre-positioning and real-time allocation of mobile emergency generator as emergency response to natural disasters. It has been shown that efficient pre-disaster planning and allocation of mobile emergency generator expedite service restoration to customers. Authors in \cite{gao2017resilience} have studied pre-hurricane planning and allocation of electric buses in distribution systems.

Moreover, utilization of pre-existing distributed energy resources along with microgrid and networked microgrids formation after natural hazards has been extensively studied in last decade. References \cite{arif2017networked, resende2011service, Moreira2007, Wang2016a, Chen2016, 10559584} have investigated microgrid and networked microgrids formation with the use of distributed energy resources after natural hazards. They have also provided sequence of control actions required for such microgrid and networked microgrids operation.

In addition, preceding this decade research activities related to power system restoration are mainly focused on transmission system restoration, treating distribution systems as passive components. References \cite{nagata2002multi, sakaguchi1983development, adibi2000power, adibi1994power, adibi1987power, adibi1991power} have performed such investigation in details. They have described restoration strategies and issues for bulk power system restoration using conventional generators and bulk transmission systems.

In last two-three years, research communities have started investigation on coordinated restoration of transmission and distribution systems. Due to the presence of large amount of distributed energy resources in distribution systems, it has been realized that distribution systems should be actively participating in restoration of post-disaster power grid. Literature \cite{zhao2019coordinated, sun2019distributed,9000318} have shepherded such research activities. As distribution systems continue to evolve in the form of smart grid, research activities which actively involve distribution systems in bulk power system restoration such as coordinated restoration of transmission and distribution systems \cite{zhao2019coordinated, sun2019distributed} should be invested and rewarded more in the future.

In summary, above-reviewed work can be broadly categorized into two: 1) work related to pre-disaster planning and scheduling of emergency response resources \cite{lei2016mobile, gao2017resilience} 2) work related to restoration of transmission and distribution systems \cite{arif2017networked, resende2011service, Moreira2007, Wang2016a, Chen2016, nagata2002multi, sakaguchi1983development, adibi2000power, adibi1994power, adibi1987power, adibi1991power, zhao2019coordinated, sun2019distributed}. It is observed that category 1) work have been performed without considering coordinated operation of transmission and distribution systems. Planning and scheduling of emergency response resources have been investigated for either distribution system or transmission systems, without the coordination of both. Work related to category 2) consider either coordinated restoration of transmission and distribution systems \cite{zhao2019coordinated, sun2019distributed} or restoration of distribution and transmission systems separately \cite{arif2017networked, resende2011service, Moreira2007, Wang2016a, Chen2016, nagata2002multi, sakaguchi1983development, adibi2000power, adibi1994power, adibi1987power, adibi1991power}. In this paper, we have proposed coordinated operation of transmission and distribution systems to determine load pickup amount using centralized generation and DERs in distribution systems. In this paper, we have shown that coordination of transmission and distribution systems expedites the load restoration process.

The rest of the paper is organized as follows: Section II provides description of proposed model and Section III describes deterministic mathematical formulation of the model. Section IV elaborates stochastic modeling of uncertain parameters in the model. Section V provides solution methodology, and Section VI provides case study and results of presented optimization model. Section VII concludes with conclusion and potential future research.

\section{Problem Description}
Normally, the load restoration is preceded by different stages such as start of black start units, start of non-black start units, and establishing transmission grid. Therefore, in load restoration, it is assumed that system is established strongly enough so that DERs in distribution systems can be connected back to the grid. In addition, before the start of load restoration stage, it is assumed that all the transmission buses and lines are energized and in operation except for the permanently damaged buses/nodes. In load restoration stage, ramping rates of generators are the limiting factors; and in this paper, using DERs from distribution systems, negative impact of ramping rates of generators is minimized. In other words, through the coordination of transmission and distribution systems, DERs in distribution systems are utilized to supply the most critical loads at the beginning, and outage time for critical loads and normal loads are minimized. With time, output of generators is increased, more loads are picked up. Therefore, due to various such benefits of utilizing DERs in load restoration, the problem of load restoration considering characteristics and operation of both transmission and distribution systems is formulated as optimization model. The following subsection describes different formulation techniques that have been applied to solve the coordination of transmission and distribution systems in load restoration.

\subsection{Problem Formulation}
The load restoration problem considering coordination of transmission and distribution system is a relatively new research field and studied in a very few papers \cite{sun2019distributed, zhao2019coordinated}. The coordination problem is mainly formulated using three different approach: integrated or centralized approach \cite{gyugyi1992unified}, distributed or decentralized approach \cite{sun2019distributed}, and hybrid approach \cite{zhao2019coordinated}. The transmission system is usually operated by transmission system operator (TSO) or regional transmission operator (RTO); and transmission system operation is normally formulated as one optimization model considering distribution system as an aggregated load at a transmission bus. Similarly, distribution system is operated by distribution system operator (DSO); and, distribution system operation is formulated as another optimization model treating transmission system as a substation node. The centralized coordination approach combines optimization models of TSO and DSOs and solve combined optimization model in a centralized fashion. The main advantages of such approach is: it is easy to implement and is the most accurate solving approach if it is solved. However, it is very difficult to solve for larger systems and computationally burdensome. Furthermore, information need to be collected at a place, therefore, it demands high communication requirements. In contrast, decentralized or distributed approach solves TSO and DSO optimization models in distributed fashion with limited exchange of information. Since only required information are exchanged between TSO and DSOs, it preserves the privacy of data. In addition, communication requirement is also relatively low due to limited communication between operators. Furthermore, TSO optimization model and DSO optimization models are solved in parallel; hence, it is computationally efficient. However, distributed algorithms such as alternating direction method of multipliers (ADMMs), bender's decomposition, optimality condition decomposition (OCD) require that TSO and DSOs optimization models in convex form, introducing errors usually. Furthermore, distributed algorithms do not guarantee optimality and convergence if both TSO optimization model and DSOs optimization models have binary or integer variables together. The hybrid approach combines the ability of centralized approach to work with binary or integer variables and the ability of distributed approach to solve in parallelism. Hybrid approach adopts two-layered hierarchical method: top layer makes decision on binary or integer variables using centralized approach and bottom layer works with continuous variables using distributed approach.

\begin{figure}[t!]
\centering
\includegraphics[scale=0.9]{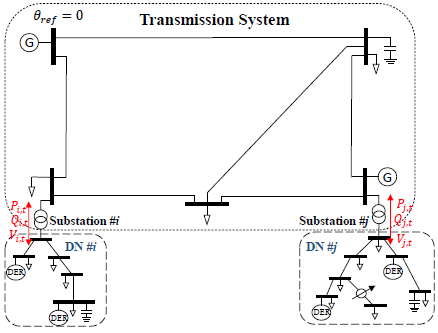}
\caption{A post-disaster power system\cite{sun2019distributed}}
\label{TDintegrated}
\end{figure}

In this paper, centralized/integrated approach is used for modeling the coordination of transmission and distribution systems in load restoration. The objective of the model is to minimize the load or energy not served in both transmission and distribution systems. In this paper, DERs are optimally and efficiently utilized to minimize the outage time of critical and non-critical loads in transmission and distribution systems. Constraints of optimization model include power flows limits, bus voltage limits, load limits (to account for critical loads), operating and technical constraints of centralized generation as well as DERs. Optimization model is built considering coordinated operation of transmission and distribution systems as shown in figure \ref{TDintegrated} where multiple distribution networks (DNs) are connected to a transmission network. In compact form, optimization model is presented as follows:
\begin{dmath} \label{eq1}
min \; f^{T}(X^{T},X_{B}^{T})+\sum_{i \in D}f^{D_{i}}(X^{D_{i}},X_{B}^{D_{i}})
\end{dmath}

\begin{dmath} \label{eq2}
h^{T}(X^{T},X_{B}^{T})=0
\end{dmath}

\begin{dmath} \label{eq3}
g^{T}(X^{T},X_{B}^{T})\leq 0 
\end{dmath}

\begin{dmath} \label{eq4}
h^{D_{i}}(X^{D_{i}},X_{B}^{D_{i}})=0
\end{dmath}

\begin{dmath} \label{eq5}
g^{D_{i}}(X^{D_{i}},X_{B}^{D_{i}})\leq 0
\end{dmath}

\begin{dmath} \label{eq6}
X_{B}^{T}=X_{B}^{D}
\end{dmath}
It is worth mentioning that functions, variables and constraints with superscripts $T$ or $D_{i}$ are only related to TSO and DSOs respectively. Transmission and distribution systems are coupled through boundary variables denoted by $X_{B}^{T}$ or  $X_{B}^{D_{i}}$. Equation (\ref{eq1}) is composed of objective function related to transmission system (first part) and distribution systems (second part). Equations (\ref{eq2}) and (\ref{eq3}) are equality and inequality constraints related to transmission systems. Similarly, equations (\ref{eq4}) and (\ref{eq5}) are equality and inequality constraints related to distribution systems. Equation (\ref{eq6}) ensures that coupling variables have same value in both TSO optimization model and DSOs optimization models. Detailed mathematical modeling of objective function and constraints related to both transmission and distribution systems is presented below.

\subsection{Objective Function}
The objective function is to minimize the energy not served in transmission and distribution systems. Mathematically, it is represented as
\begin{dmath} \label{Ob1}
min \sum_{t}\left (W^{T} \sum_{i}\left ( P_{i,t}^{T,load\_total} -P_{i,t}^{T,load}\right ) + \sum_{D_{i}} W^{D_{i}}\sum_{i} \left ( P_{i,t}^{D_{i},load\_total} -P_{i,t}^{D_{i},load} \right ) \right ),
\end{dmath}
where $P_{i,t}^{T,load\_total}$ $\&$ $P_{i,t}^{T,load}$ are total and supplied load of bus $i$ at time $t$ in TN, and $P_{i,t}^{D_{i},load\_total}$ $\&$ $P_{i,t}^{D_{i},load}$ are total and supplied load of node $i$ at time $t$ in DN $D_{i}$. In addition, $W^{T}$ $\&$ $W^{D_{i}}$ are weight factors for transmission and distribution networks load respectively.

\subsection{Constraints}
\subsubsection{Power flow constraints}
Full non-linear original ac power flow (ACPF) equations are used to calculate power flows between any two buses $i$ and $k$ in transmission system and DISTFLOW model based power flow equations are used to calculate power flows between any two nodes $i$ and $k$ in distribution systems. For further details, please see \cite{gyugyi1992unified}. 

\subsubsection{Power balance constraints}
Power balance equations (\ref{Pb1}-\ref{Pb4}) ensure active and reactive power balance at each node in both transmission and distribution networks. Right hand side of equations (\ref{Pb1}-\ref{Pb4}) represents difference of power generation and power demand at a node. Similarly, left hand side of equations (\ref{Pb1}-\ref{Pb4}) represents difference of incoming power flow and outgoing power flow at the same node. 
\begin{dmath} \label{Pb1}
P_{substation\_node,t}^{D_{i},grid}+P_{i,t}^{D_{i},DG}+P_{i,t}^{D_{i},ESS}+P_{i,t}^{D_{i},PV}-P_{i,t}^{D_{i},load}= \sum_{ji}P^{D_{i}}_{ji,t}-\sum_{ik}P^{D_{i}}_{ik,t},
\end{dmath}
where $P$ refers to active power generation, superscripts $DG$, $ESS$, and $PV$ refer to diesel generator, energy storage system, and solar photo-voltaic respectively. $P_{i,t}^{D_{i},DG}$ is read as active power generated by diesel generator at node $i$ at time $t$ in distribution system $D_{i}$.

\begin{dmath} \label{Pb2}
Q_{substation\_node,t}^{D_{i},grid}+Q_{i,t}^{D_{i},DG}+Q_{i,t}^{D_{i},ESS}+Q_{i,t}^{D_{i},PV}-Q_{i,t}^{D_{i},load}= \sum_{ji}Q^{D_{i}}_{ji,t}-\sum_{ik}Q^{D_{i}}_{ik,t},
\end{dmath}
where, $Q$ refers to reactive power generation.
\begin{dmath} \label{Pb3}
P_{i,t}^{T,G}-P_{i,t}^{T,load}= \sum_{ji}P^{T}_{ji,t}-\sum_{ik}P^{T}_{ik,t},
\end{dmath}
where superscript $G$ refers to centralized generation in transmission system, and $P_{i,t}^{T,G}$ is read as active power generation at bus $i$ in transmission system.
\begin{equation} \label{Pb4}
Q_{i,t}^{T,G}-Q_{i,t}^{T,load}= \sum_{ji}Q^{T}_{ji,t}-\sum_{ik}Q^{T}_{ik,t},
\end{equation}
It is worth mentioning that line flows are calculated using non-linear ac power flow equations in both transmission and distribution systems \cite{gyugyi1992unified}.

\subsubsection{Centralized generation constraints}
Constraints (\ref{Gen1}-\ref{Gen2}) are used to define operating characteristics of generation units in transmission.
\begin{equation} \label{Gen1}
P_{i}^{T,G,min}\leq P_{i,t}^{T,G}\leq P_{i}^{T,G,max},
\end{equation}
where it should be noted that superscripts $min$ and $max$ represent minimum and maximum limits.
\begin{equation} \label{Gen2}
Q_{i}^{T,G,min}\leq Q_{i,t}^{T,G}\leq Q_{i}^{T,G,max}
\end{equation}

\subsubsection{DG constraints}
Constraints (\ref{Gen3}-\ref{Gen4}) are used to define operating characteristics of diesel generator units in distribution networks.
\begin{equation} \label{Gen3}
P_{i}^{D_{i},DG,min}\leq P_{i,t}^{D_{i},DG}\leq P_{i}^{D_{i},DG,max}
\end{equation}
\begin{equation} \label{Gen4}
Q_{i}^{D_{i},DG,min}\leq Q_{i,t}^{D_{i},DG}\leq Q_{i}^{D_{i},DG,max}
\end{equation}

\subsubsection{ESS constraints}
Constraints (\ref{es3}-\ref{es5}) are used to model operating characteristics of energy storage systems in distribution networks.
\begin{dmath} \label{es3}
0\leq E_{i}^{D_{i},ESS\_spl}-\sum_{t}\left ( P_{i,t}^{D_{i},ESS}+P_{i,t}^{D_{i},ESS\_loss} \right )\Delta t \leq
E_{i}^{D_{i},ESS,max},
\end{dmath}
where $E_{i}^{D_{i},ESS\_spl}$ and $E_{i}^{D_{i},ESS,max}$ represent surplus energy and maximum energy capacity of energy storage system at node $i$ in distribution system $D_{i}$. Furthermore, $P_{i,t}^{D_{i},ESS\_loss}$ refers to power loss in energy storage at node $i$ in distribution system $D_{i}$.
\begin{equation} \label{es4}
\left ( P_{i,t}^{D_{i},ESS} \right )^2+\left ( Q_{i,t}^{D_{i},ESS} \right )^2\leq \left ( S_{i}^{D_{i},ESS,max} \right )^2,
\end{equation}
where $S_{i}^{D_{i},ESS,max}$ represents MVA capacity of energy storage system at node $i$ in distribution system $D_{i}$.
\begin{dmath} \label{es5}
rEq_{i}^{D_{i},ESS}\left ( P_{i,t}^{D_{i},ESS} \right )^2+rCvt_{i}^{D_{i},ESS}\left ( Q_{i,t}^{D_{i},ESS} \right )^2= P_{i,t}^{D_{i},ESS\_loss}V^{D_{i}}_{i,t},
\end{dmath}
where $rEq_{i}^{D_{i},ESS}$ and $rCvt_{i}^{D_{i},ESS}$ represent resistance of energy storage and converter. In addition, $V^{D_{i}}_{i,t}$ represents squared of voltage at node $i$ in distribution system $D_{i}$.

\subsubsection{Solar PV constraints}
Following constraints are used to model the technical and operating characteristics of solar PV systems.
\begin{equation} \label{pv1}
0\leq P_{i,t}^{D_{i},PV}\leq P_{i}^{D_{i},PV,max}Profile^{D_{i}}_{t},
\end{equation}
where $P_{i}^{D_{i},PV,max}$ refers to maximum capacity of PV at node $i$ in distribution system $D_{i}$, $Profile^{D_{i}}_{t}$ refers to PV profile at time $t$ in distribution system $D_{i}$. PVs are designed to operate at fixed power factor $PF$, reactive power output from PVs are modeled by following constraint.
\begin{equation} \label{pv2}
-PF^{D_{i}}P_{i,t}^{D_{i},PV}\leq Q_{i,t}^{D_{i},PV}\leq PF^{D_{i}}P_{i,t}^{D_{i},PV}
\end{equation}

\subsubsection{Load limits}
Constraints (\ref{ld1}-\ref{ld4}) define load limits in both transmission and distribution networks.
\begin{equation} \label{ld1}
P_{i}^{T,load\_critical}\leq P_{i,t}^{T,load}\leq P_{i}^{T,load\_total}
\end{equation}

\begin{equation} \label{ld2}
Q_{i}^{D_{i},load\_critical}\leq Q_{i,t}^{D_{i},load}\leq Q_{i}^{D_{i},load\_total}
\end{equation}

\begin{equation} \label{ld3}
P_{i}^{D_{i},load\_critical}\leq P_{i,t}^{D_{i},load}\leq P_{i}^{D_{i},load\_total}
\end{equation}

\begin{equation} \label{ld4}
Q_{i}^{T,load\_critical}\leq Q_{i,t}^{T,load}\leq Q_{i}^{T,load\_total}
\end{equation}
It should be noted that superscript $load\_critical$ refers to critical load.

\subsubsection{Voltage limits}
Constraints (\ref{vl1}-\ref{vl2}) represent voltage limits in both transmission and distribution networks.
\begin{equation} \label{vl1}
V_{i}^{T,min}\leq V_{i,t}^{T}\leq V_{i}^{T,max}
\end{equation}

\begin{equation} \label{vl2}
V_{i}^{D,min}\leq V_{i,t}^{D}\leq V_{i}^{D,max}
\end{equation}
The voltage magnitude is constrained between 0.95 pu and 1.05 pu most of the time in both transmission and distribution systems.

\subsubsection{Boundary constraints}
Boundary constraints in (\ref{eq5}) ensure that variables that couple transmission and distribution systems have equal values. Coupling variables are active and reactive power exchange, and voltage at substation bus/node. Consider an example where node 1 of a $DN_{i}$ is connected to bus 1 of a transmission network.  Mathematically, boundary constraints are represented as follows for aforementioned example
\begin{equation} \label{bc1}
P_{1,t}^{T,load}=P_{1,t}^{D_{i},grid},
\end{equation}
where $P_{1,t}^{D_{i},grid}$ refers to active power generation from transmission system in distribution system $D_{i}$.
\begin{equation} \label{bc2}
Q_{1,t}^{T,load}=Q_{1,t}^{D_{i},grid}
\end{equation}
\begin{equation} \label{bc3}
V_{1,t}^{T}=V_{1,t}^{D_{i},grid}.
\end{equation}
It is worth mentioning that when transmission system is feeding distribution systems, power exchange is treated as load in transmission system and as generation in distribution systems, which has been reflected in equations (\ref{bc1}-\ref{bc3}).

\begin{figure}[H]
\centering
\includegraphics[scale=0.3]{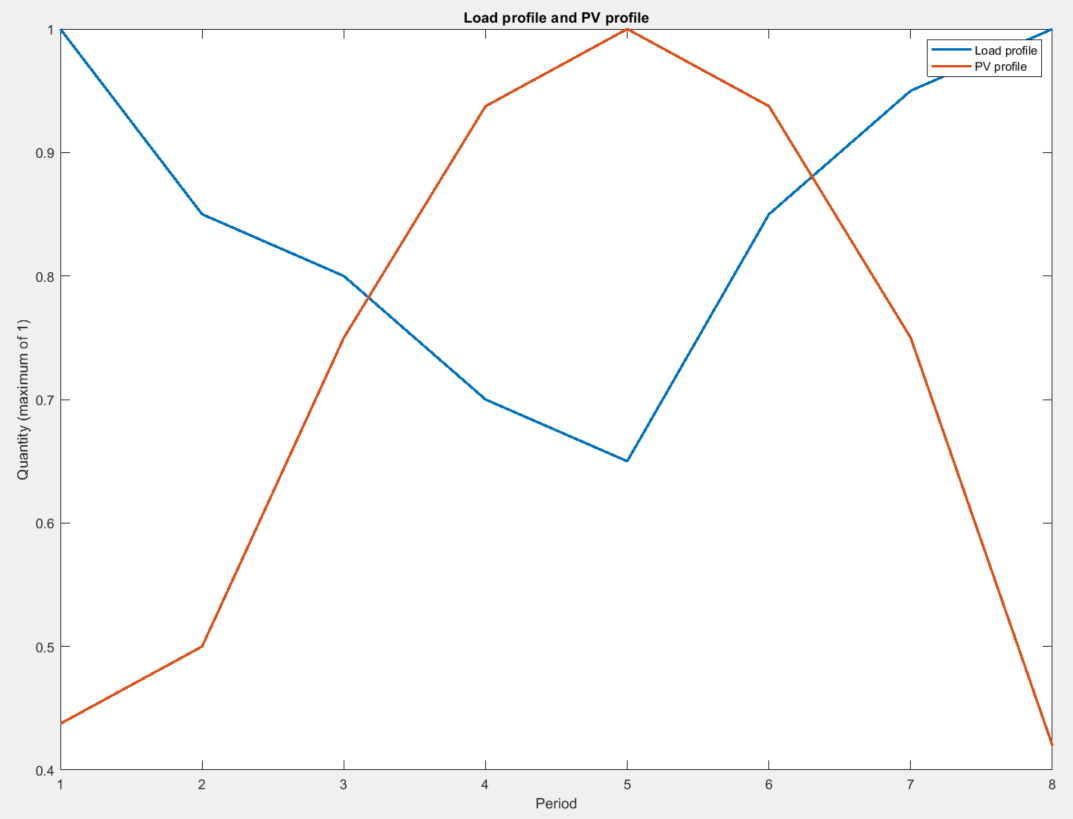}
\caption{Load and PV profiles}
\label{Profile}
\end{figure}

\begin{table*}[t!]
\begin{center}
\caption{DERs data for distribution networks.}
\label{tab:my-table}
\resizebox{0.75\textwidth}{!}{%
\begin{tabular}{|c|c|c|c|c|}
\hline
\begin{tabular}[c]{@{}c@{}}Distribution \\ system\end{tabular} & \begin{tabular}[c]{@{}c@{}}Transmission \\ bus\end{tabular} & DGs                                                                                             & ESSs                                                                       & PVs                                                             \\ \hline
$D_{1}$                                                           & 5                                                           & \begin{tabular}[c]{@{}c@{}}2 units of 4MW/3.2MVAr each \\ DGs at node 1 and node 8\end{tabular} & \begin{tabular}[c]{@{}c@{}}50MWh/25MVA\\ of ESS at node 3\end{tabular}     & \begin{tabular}[c]{@{}c@{}}3MW of PV \\ at node 11\end{tabular} \\ \hline
$D_{2}$                                                            & 9                                                           & \begin{tabular}[c]{@{}c@{}}2 units of 1MW/0.8MVAr each \\ DGs at node 1 and node 8\end{tabular} & \begin{tabular}[c]{@{}c@{}}12.5MWh/6.25MVA\\ of ESS at node 3\end{tabular} & \begin{tabular}[c]{@{}c@{}}3MW of PV \\ at node 11\end{tabular} \\ \hline
$D_{3}$                                                           & 14                                                          & \begin{tabular}[c]{@{}c@{}}2 units of 14MW/9MVAr each \\ DGs at node 1 and node 8\end{tabular}  & \begin{tabular}[c]{@{}c@{}}50MWh/25MVA \\ of ESS at node 3\end{tabular}    & \begin{tabular}[c]{@{}c@{}}3MW of PV \\ at node 11\end{tabular} \\ \hline
\end{tabular}%
}
\end{center}
\end{table*}

\begin{table*}[htbp]
\caption{Transmission system loads data}
\label{tab:my-table1}
\resizebox{\textwidth}{!}{%
\begin{tabular}{|c|c|c|c|c|c|c|c|c|c|c|c|c|c|l|}
\hline
Bus & 1 & 2 & 3 & 4 & 5 & 6 & 7 & 8 & 9 & 10 & 11 & 12 & 13 & 14 \\ \hline
\begin{tabular}[c]{@{}c@{}}Active power\\ (MW)\end{tabular} & 0 & 21.70 & 94.20 & 47.80 & 10.398 & 11.20 & 0 & 0 & 34.66 & 9 & 3.50 & 6.10 & 13.50 & 17.33 \\ \hline
\begin{tabular}[c]{@{}c@{}}Reactive power\\ (MVAr)\end{tabular} & 0 & 12.70 & 19 & -3.90 & 5.0448 & 7.50 & 0 & 0 & 16.816 & 5.80 & 1.80 & 1.60 & 5.80 & 8.408 \\ \hline
\end{tabular}%
}
\end{table*}

\begin{table*}[htbp]
\caption{distribution system loads data}
\label{tab:my-table2}
\resizebox{\textwidth}{!}{%
\begin{tabular}{|l|l|l|l|l|l|l|l|l|l|l|l|l|l|l|}
\hline
 & \multicolumn{1}{c|}{Node} & \multicolumn{1}{c|}{1} & \multicolumn{1}{c|}{2} & \multicolumn{1}{c|}{3} & \multicolumn{1}{c|}{4} & \multicolumn{1}{c|}{5} & \multicolumn{1}{c|}{6} & \multicolumn{1}{c|}{7} & \multicolumn{1}{c|}{8} & \multicolumn{1}{c|}{9} & \multicolumn{1}{c|}{10} & \multicolumn{1}{c|}{11} & \multicolumn{1}{c|}{12} & \multicolumn{1}{c|}{13} \\ \hline
\multirow{2}{*}{$D_{1}$} & \multicolumn{1}{c|}{\begin{tabular}[c]{@{}c@{}}Active power\\ (MW)\end{tabular}} & \multicolumn{1}{c|}{0.51} & \multicolumn{1}{c|}{0.30} & \multicolumn{1}{c|}{0} & \multicolumn{1}{c|}{1.20} & \multicolumn{1}{c|}{0.51} & \multicolumn{1}{c|}{0.69} & \multicolumn{1}{c|}{0} & \multicolumn{1}{c|}{0.384} & \multicolumn{1}{c|}{3.765} & \multicolumn{1}{c|}{2.529} & \multicolumn{1}{c|}{0} & \multicolumn{1}{c|}{0} & \multicolumn{1}{c|}{0.51} \\ \cline{2-15} 
 & \multicolumn{1}{c|}{\begin{tabular}[c]{@{}c@{}}Reactive power\\ (MVAr)\end{tabular}} & \multicolumn{1}{c|}{0.1920} & \multicolumn{1}{c|}{0.1392} & \multicolumn{1}{c|}{0} & \multicolumn{1}{c|}{0.6960} & \multicolumn{1}{c|}{0.3000} & \multicolumn{1}{c|}{0.3168} & \multicolumn{1}{c|}{0} & \multicolumn{1}{c|}{0.2064} & \multicolumn{1}{c|}{1.7232} & \multicolumn{1}{c|}{1.1088} & \multicolumn{1}{c|}{0} & \multicolumn{1}{c|}{0} & \multicolumn{1}{c|}{0.3624} \\ \hline
\multirow{2}{*}{$D_{2}$} & \begin{tabular}[c]{@{}l@{}}Active power\\ (MW)\end{tabular} & 1.70 & 1.00 & 0 & 4.0 & 1.70 & 2.30 & 0 & 1.28 & 12.55 & 8.43 & 0 & 0 & 1.70 \\ \cline{2-15} 
 & \begin{tabular}[c]{@{}l@{}}Reactive power\\ (MVAr)\end{tabular} & 0.64 & 0.464 & 0 & 2.32 & 1.00 & 1.056 & 0 & 0.688 & 5.744 & 3.696 & 0 & 0 & 1.208 \\ \hline
\multirow{2}{*}{$D_{3}$} & \begin{tabular}[c]{@{}l@{}}Active power\\ (MW)\end{tabular} & 0.85 & 0.50 & 0 & 2.00 & 0.85 & 1.15 & 0 & 0.64 & 6.275 & 4.215 & 0 & 0 & 0.85 \\ \cline{2-15} 
 & \begin{tabular}[c]{@{}l@{}}Reactive power\\ (MVAr)\end{tabular} & 0.32 & 0.232 & 0 & 1.16 & 0.50 & 0.528 & 0 & 0.344 & 2.872 & 1.848 & 0 & 0 & 0.604 \\ \hline
\end{tabular}%
}
\end{table*}

\section{Case Study and Results}
The load restoration problem considering coordination of transmission and distribution systems is formulated as centralized/integrated non-linear optimization model in Section II. The non-linearities are introduced by non-linear power flow models, and non-linear modeling of energy storage systems. The IEEE 14 bus test case is used as transmission system, and three modified IEEE 13 node test feeders are used as three different distribution systems and are connected to bus 5, bus 9, and bus 14, respectively, of transmission system. The loads in bus 5, bus 9, and bus 14 of original IEEE 14 test system are replaced by distribution systems, loads on rest of transmission buses are treated as transmission loads. DGs, ESSs, and PVs are connected to all three distribution systems, exact locations and amounts are provided in Table I. From the table, it is inferred that distribution systems $D_{1}$ and $D_{3}$ have relatively high level of DERs than distribution system $D_{2}$. The total loads of all the buses in transmission system and loads of all the nodes in distribution systems are provided in Table II and III respectively. From the tables II and III, it is noted that total loads of distribution network connected to transmission bus 5 is equal to transmission load assigned to that bus, similar for other buses 9 and 14. From the tables I, II, and III, it is inferred that distribution system-2 has relatively high loads and low DERs levels than other two distribution networks. Since load restoration is designed as the multi-period operation problem, load profile and PV profile are used, load profile and PV profile are as shown in figure 2. Other parameters such as line resistance, line reactance are obtained from MATPOWER \cite{zimmerman2010matpower} and reference \cite{kersting2012short}. The optimization model is solved using YALMIP tool of MATLAB using the solver FMINCON \cite{lofberg2004yalmip}. Following subsections provide two case study of the load restoration problem. Case studies focus mainly on how to optimize the output of DERs in load restoration stage. For both case study, convergence or consensus on boundary variables is achieved using integrated of centralized coordination strategy described in Section II.

\subsection{Case Study I}
The transmission generation for this case study is provided in Table IV. All the loads, both transmission and distribution systems, have equal priority. The critical load is assumed to be 50\% of total loads in all nodes and buses. From the Table IV, it is seen that centralized generation available is very high, sufficient enough to supply all the loads by itself, without needing DERs. Therefore, in order to utilize the DERs, one additional term is introduced in the objective function which tries to keep the utilization of centralized generation as minimum as possible. For every 1 MW of centralized generation, objective is increased by 10000000. With the introduction of new term in the objective function, optimization model seeks to maximize the utilization of DERs and minimize the use of transmission generation. The values of boundary variables for period-3 are provided in Table V, due to the limitation of the space, values of boundary variables for rest of the periods are omitted from the report. The output of transmission generation is provided in Table VI. The total load picked up in this case study is 50\% of total loads in both transmission and distribution systems. From the Table V, it is inferred that distribution systems 1 and 3 supply power to the grid and distribution system 2 gets power from the grid. This is because distribution systems 1 and 3 have relatively high level of DERs but low power demand; however, distribution system 2 has relatively low level of DERs but high power demand. It is worth mentioning that negative value of power exchange for a distribution system indicate that the distribution system is supplying power to the grid. Similarly, positive value of power exchange for a distribution system indicates that the distribution system is getting power from the grid. Since the cost associated with utilizing centralized generation is very high, distribution systems supply their critical loads only, remaining available power they feed to grid so that critical loads of transmission system are picked utilizing as minimum of centralized generation as feasible, as shown in Table VI. 

\begin{table}[H]
\caption{Transmission generation parameters for Case Study I.}
\label{tab:my-table3}
\resizebox{\linewidth}{!}{%
\begin{tabular}{|c|c|c|c|c|}
\hline
Bus & \begin{tabular}[c]{@{}c@{}}$P^{T,G,max}$\\ (MW)\end{tabular} & \begin{tabular}[c]{@{}c@{}}$P^{T,G,min}$\\ (MW)\end{tabular} & \begin{tabular}[c]{@{}c@{}}$Q^{T,G,max}$\\ (MVAr)\end{tabular} & \begin{tabular}[c]{@{}c@{}}$Q^{T,G,min}$\\ (MVAr)\end{tabular} \\ \hline
1 & 332.4 & 0 & 10 & 0 \\ \hline
2 & 140 & 0 & 50 & -40 \\ \hline
3 & 0 & 0 & 40 & 0 \\ \hline
6 & 0 & 0 & 24 & -6 \\ \hline
8 & 0 & 0 & 24 & -6 \\ \hline
\end{tabular}%
}
\end{table}

\begin{table}[H]
\caption{Boundary variables for Case Study I.}
\label{tab:my-table5}
\resizebox{\linewidth}{!}{%
\begin{tabular}{|c|c|c|c|}
\hline
\begin{tabular}[c]{@{}c@{}}Distribution\\ System\end{tabular} & \begin{tabular}[c]{@{}c@{}}Active power\\  exchange (MW)\end{tabular} & \begin{tabular}[c]{@{}c@{}}Reactive power \\ exchange (MVAr)\end{tabular} & Voltage (pu) \\ \hline
$D_{1}$ & -6.96 & -6.11 & 1.0549 \\ \hline
$D_{2}$ & 9.74 & 5.24 & 1.0402 \\ \hline
$D_{3}$ & -25.46 & 0.66 & 1.0556 \\ \hline
\end{tabular}%
}
\end{table}

\begin{table}[H]
\begin{center}
\caption{Transmission generation output for Case Study I.}
\label{tab:my-table6}
\resizebox{0.75\linewidth}{!}{%
\begin{tabular}{|c|c|c|c|c|c|}
\hline
Bus & 1 & 2 & 3 & 6 & 8 \\ \hline
$P^{T,G}$ & 7.14 & 53.99 & 0 & 0 & 0 \\ \hline
$Q^{T,G}$ & 0.69 & 9.28 & 14.15 & 0.47 & 3.71 \\ \hline
\end{tabular}%
}
\end{center}
\end{table}

\subsection{Case Study II}
In this case study, the distribution systems loads have equal priority of 1, and all the transmission loads have equal priority of 2 (higher priority). Therefore, not serving transmission loads increases the objective function by 2 times than not serving the same amount of distribution system loads. The critical load is assumed to be 50\% of total loads in all nodes and buses. The transmission generation for this case study is provided in Table VII.  From the Table VII, it is seen that centralized generation available is very high, sufficient enough to supply all the loads by itself, without needing DERs. Therefore, in order to utilize the DERs, one additional term is introduced in the objective function which tries to keep the utilization of centralized generation as minimum as possible. For every 1 MW of centralized generation, objective is increased by 10000000. With the introduction of new term in the objective function, optimization model seeks to maximize the utilization of DERs and minimize the use of transmission generation. The values of boundary variables for period-3 are provided in Table VIII, due to the limitation of the space, values of boundary variables for rest of the periods are omitted from the report. The output of transmission generation is provided in Table IX. From the Table IX, it is seen that generator at bus 2 in transmission system is operating at its upper limit; therefore, generator at bus 1 in transmission system is supplying more power than in Case Study I. The total load picked up in this case study is 50\% of total loads in both transmission and distribution systems. From the Table V, it is inferred that distribution systems 1 and 3 supply power to the grid and distribution system 2 gets power from the grid. This is because distribution systems 1 and 3 have relatively high level of DERs but low power demand; however, distribution system 2 has relatively low level of DERs but high power demand.  Since the cost associated with utilizing centralized generation is very high, distribution systems supply their critical loads only, remaining available power they feed to grid so that critical loads of transmission system are picked utilizing as minimum of centralized generation as feasible, as shown in Table IX. 

\begin{table}[H]
\caption{Transmission generation parameters for Case Study II.}
\label{tab:my-table9}
\resizebox{\linewidth}{!}{%
\begin{tabular}{|c|c|c|c|c|}
\hline
Bus & \begin{tabular}[c]{@{}c@{}}$P^{T,G,max}$\\ (MW)\end{tabular} & \begin{tabular}[c]{@{}c@{}}$P^{T,G,min}$\\ (MW)\end{tabular} & \begin{tabular}[c]{@{}c@{}}$Q^{T,G,max}$\\ (MVAr)\end{tabular} & \begin{tabular}[c]{@{}c@{}}$Q^{T,G,min}$\\ (MVAr)\end{tabular} \\ \hline
1 & 114.62 & 0 & 5.26 & 0 \\ \hline
2 & 48.28 & 0 & 26.32 & -40 \\ \hline
3 & 0 & 0 & 21.05 & 0 \\ \hline
6 & 0 & 0 & 12.63 & -6 \\ \hline
8 & 0 & 0 & 12.63 & -6 \\ \hline
\end{tabular}%
}
\end{table}

\begin{table}[H]
\caption{Boundary variables for Case Study II.}
\label{tab:my-table11}
\resizebox{\linewidth}{!}{%
\begin{tabular}{|c|c|c|c|}
\hline
\begin{tabular}[c]{@{}c@{}}Distribution\\ System\end{tabular} & \begin{tabular}[c]{@{}c@{}}Active power\\  exchange (MW)\end{tabular} & \begin{tabular}[c]{@{}c@{}}Reactive power \\ exchange (MVAr)\end{tabular} & Voltage (pu) \\ \hline
$D_{1}$ & -5.93 & -5.85 & 1.0543 \\ \hline
$D_{2}$ & 9.62 & 4.99 & 1.0401 \\ \hline
$D_{3}$ & -25.35 & 0.59 & 1.0555 \\ \hline
\end{tabular}%
}
\end{table}

\begin{table}[H]
\begin{center}
\caption{Transmission generation output for Case Study II.}
\label{tab:my-table16}
\resizebox{0.75\linewidth}{!}{%
\begin{tabular}{|c|c|c|c|c|c|}
\hline
Bus & 1 & 2 & 3 & 6 & 8 \\ \hline
$P^{T,G}$ & 13.87 & 48.28 & 0 & 0 & 0 \\ \hline
$Q^{T,G}$ & 0.32 & 9.85 & 12.93 & 1.86 & 3.62 \\ \hline
\end{tabular}%
}
\end{center}
\end{table}

\section{Conclusion}
DERs play pivotal role in the operation of power systems in the future due to the proliferation DERs and modernization of power grids. Power system restoration has to adopt DERs because of their several attractive features. In this project, DERs are used to support load restoration in power systems after the main skeleton of the grid is fully established. The results show that DERs can be fully utilized to restore loads and minimize the outage time for the critical and non-critical loads in both transmission and distribution systems.

\section*{Acknowledgment}
This was a result of a class project of one of author's grad courses.






\bibliographystyle{IEEEtran}
\bibliography{Project.bib}

\end{document}